\documentclass[journal]{IEEEtranTIE}
\usepackage{graphicx}
\usepackage{cite}
\usepackage{picinpar}
\usepackage{amsmath}
\usepackage{mathtools} 
\usepackage{url}
\usepackage{flushend}
\usepackage[latin1]{inputenc}
\usepackage{colortbl}
\usepackage{soul}
\usepackage{multirow}
\usepackage{pifont}
\usepackage{color}
\usepackage{alltt}
\usepackage[hidelinks]{hyperref}
\usepackage{enumerate}
\usepackage{siunitx}
\usepackage{breakurl}
\usepackage{epstopdf}
\usepackage{pbox}
\usepackage{fixltx2e} 
\usepackage{caption} 
\usepackage{subcaption} 
\graphicspath{{Figures/}} 
\begin{document}
\title{The Optimal Location and Size of an Intermediate Coil in a Magnetic Resonant Coupling Wireless Power Transfer System}

\author{
	\vskip 1em
	
	Kedi Yan, \emph{Student  Member, IEEE},
	Gregory E. Moore, \emph{Student Member, IEEE},\\
	Joshua R, Smith, \emph{Fellow, IEEE}

	
		
		
}

\maketitle

\begin{abstract}
To increase the transmission distance of Wireless Power Transfer (WPT) systems, we provide guidelines on choosing the optimal location of an Intermediate Coil with respect to size within a standard five-coil axially aligned experimental setup. From our results, for maximum magnitude of S\textsubscript{21} at the resonant frequency we found the optimal location to exist where the coupling coefficient between the Transmitter and the Intermediate Coil and the coupling coefficient between the Receiver and the Intermediate Coil are identical. Additionally, the optimal outer diameter for the maximum magnitude of S\textsubscript{21} at the resonant frequency of the Intermediate Coil in the given symmetric and asymmetric setup are found to be larger than both TX and RX. 
\end{abstract}

\begin{IEEEkeywords}
Intermediate Coil, Magnetic Resonant Coupling, Optimal Intermediate Coil Size, Optimal Intermediate Coil Location, Wireless Power Transfer
\end{IEEEkeywords}

\markboth{IEEE TRANSACTIONS ON INDUSTRIAL ELECTRONICS}
{}

\definecolor{limegreen}{rgb}{0.2, 0.8, 0.2}
\definecolor{forestgreen}{rgb}{0.13, 0.55, 0.13}
\definecolor{greenhtml}{rgb}{0.0, 0.5, 0.0}

\section{Introduction}

\IEEEPARstart{I}{n} recent years, studies in WPT have been focused on its application in the near field, such as close-range WPT that adopts the magnetic inductive coupling \cite{article1,article2,inproceedings1,inproceedings2} or mid-range WPT which utilizes Magnetic Resonant Coupling (MRC) \cite{article3,article4,article5,article6,article7,article14}. Applications range from the supplying power to commuter busses~\cite{ko_2013} all the way down to human implantable devices~\cite{borton_2013} and across many formerly hostile environments such as ocean going~\cite{hayslett_2016} or extraterrestrial based operations~\cite{liu_2015}.

In this paper, the MRC WPT is utilized, such that both transmitter coil (TX), consisting of a loop coil (TXL) and a spiral coil (TXC), and receiver coil (RX), consisting of a loop coil (RXL) and a spiral coil (RXC), are tuned to the same resonant frequency, 13.56MHz. High-power transmission of a MRC WPT system is generally realized within two coil diameters \cite{article13}, as power transfer efficiency decreases with distance, which does not still satisfy the desires of the consumer market since the diameter of the coils are generally in mm-scale. The effects of Intermediate Coils (ICs) have been shown as a means to increase the transmission distance \cite{article8,article9}, and alternative designs of such a system consisting of ICs has been proposed \cite{article10, article11}. While previous works have shown that the proper addition of one or more ICs could increase transmission distance, power transfer efficiency, and the transferred power to a conventional four-coil system, for maximum transferred power the optimal location and size for an IC are not covered.

\begin{figure}[!t]\centering
	\includegraphics[width=8.5cm]{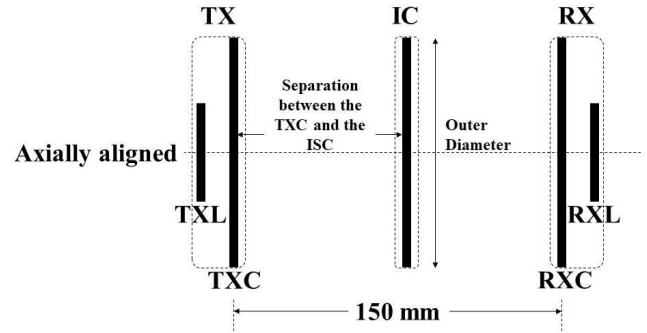}
	\caption{An Aerial view of the axially aligned five-coil system. In the examined symmetric system, the outer diameter (OD) of TX and RX are identical. In the examined asymmetric system, the OD of the TX is 3x the OD of the RX. The location of the IC, as well as the OD of the IC, are swept to determine the optimal location and size for maximum transferred power.}
	\label{fig_five_coil_system}
\end{figure}

In this paper, we will show the optimal location and size of an IC that gives the maximum transferred power in the MRC WPT system. The five-coil system used here, i.e. TXL-TXC-IC-RXC-RXL (as shown in \mbox{Fig. \ref{fig_five_coil_system}}), is analyzed by using circuit theory in Section \ref{system_analysis}. To determine the parameters that maximize the power transferred to the load, the magnitude of scattering parameter S\textsubscript{21} is quantitatively simulated and experimentally measured. Because multiple resonant frequencies will appear when coils are over coupled (so called frequency splitting in \cite{article5}), we chose to collect the S\textsubscript{21} data at the designed resonant frequency, since such frequency will provide the maximum magnitude of S\textsubscript{21} in the system of odd number of coils \cite{article9}. A symmetric and an asymmetric system have been designed, the former consists of identical TX and RX while the latter consists of a TX OD that is 3x larger than the RX OD. The system setups, including the experimental setup and quantitative simulation setup, are demonstrated in Section \ref{system_setup}. Then, in Section \ref{experimental_and_simulation_results}, the optimal location of ICs of varying sizes in both systems are provided, and simulation and experimental results are compared.

This paper provides guidance for the optimal location of an IC in the five-coil MRC WPT system, whether symmetric or asymmetric. Additionally to the optimal location, this paper reports the optimal size of the IC with the given experimental setup.  

\begin{figure}[!t]\centering
    \begin{subfigure}{0.25\textwidth}\centering
        \includegraphics[width=\textwidth]{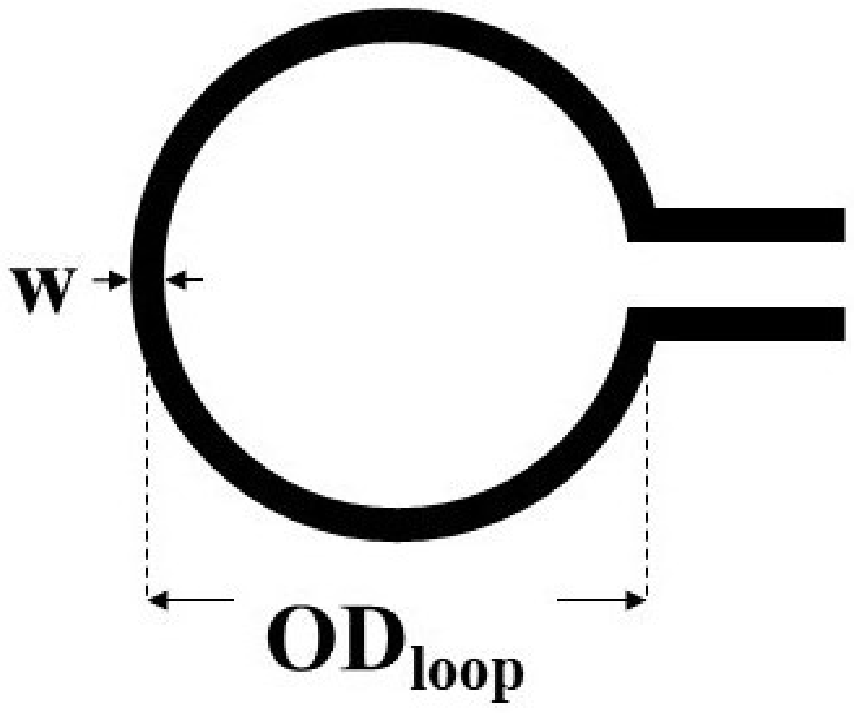}
        \caption{A loop coil}
        \label{fig_loop}
    \end{subfigure}
    \hfill
    \begin{subfigure}{0.4\textwidth}\centering
        \includegraphics[width=\textwidth]{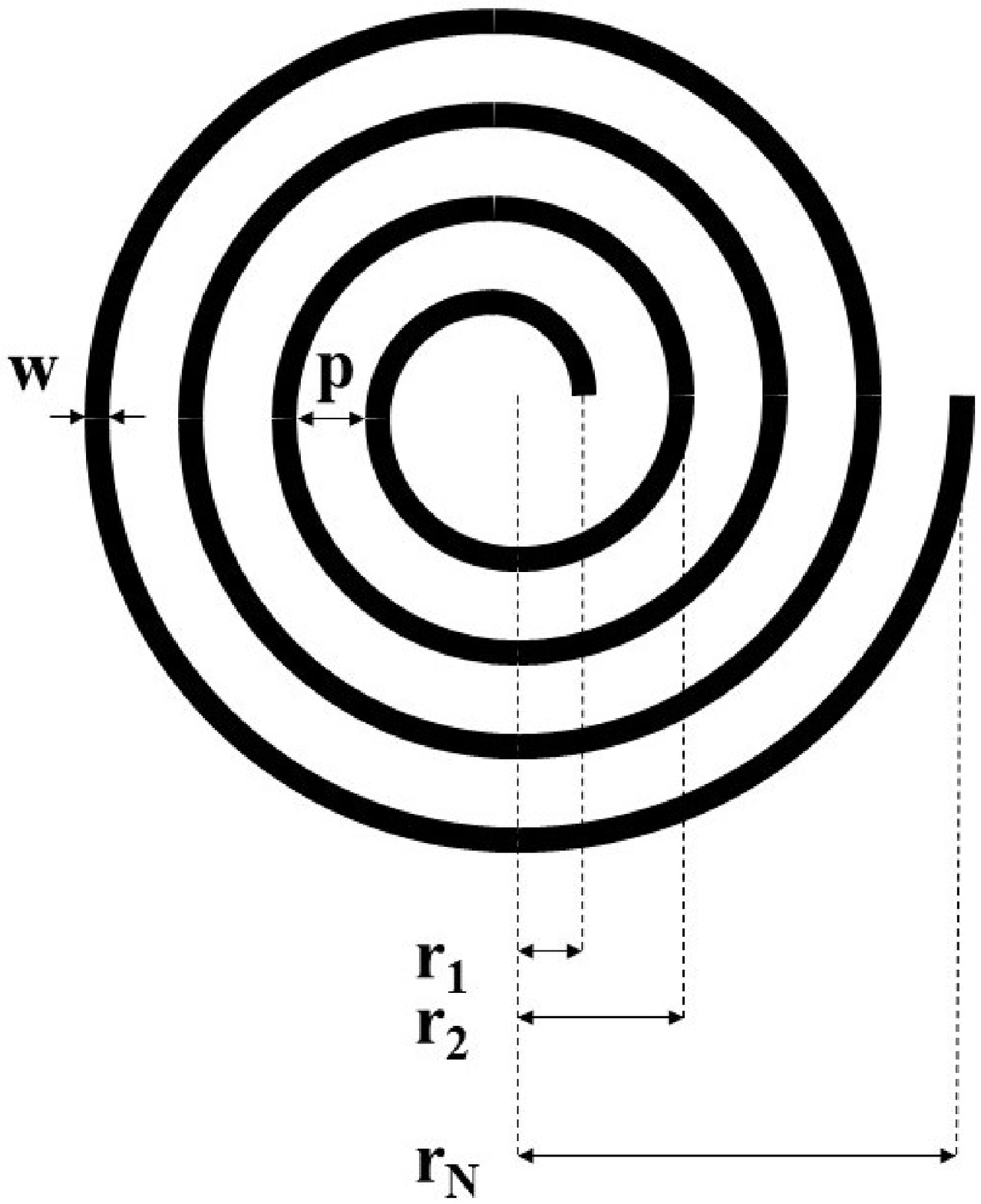}
        \caption{A spiral coil}
        \label{fig_spiral}
    \end{subfigure}
    \caption{(a) shows the geometry of a loop coil that has a OD of OD\textsubscript{loop}. (b) is the geometry of a multi-turn single layer spiral coil that has N turns, separation of p between each turn, and the radius of each turn r\textsubscript{i\textsuperscript{th}}. w is the wire diameter for both coils.}
    \label{fig_coilgeometry}
\end{figure}

\section{System Analysis} \label{system_analysis}
\begin{figure}[!t]\centering
	\includegraphics[width=8.5cm]{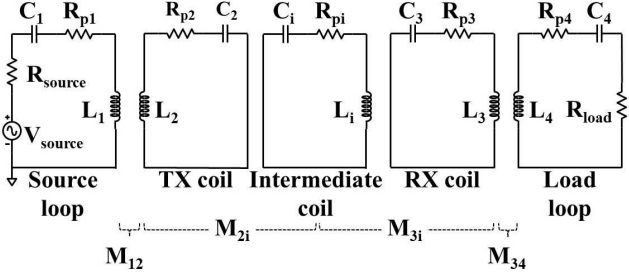}
	\caption{The equivalent circuit model of the conventional MRC WPT system with an included IC.}
	\label{fig_circuit_model}
\end{figure}

With the presence of an IC, the conventional four-coil system (loop-coil-coil-loop) becomes a five-coil system (loop-coil-intermediate-coil-loop). Then, the system is analyzed using the circuit theory so that the quantitative model can be compared with the experiment and determine the generalizability of the system. In \mbox{Fig. \ref{fig_circuit_model}}, the system circuit model is shown and all parasitic or designed components are indicated by lumped elements R, L, and C. The Source loop (TXL) and Load loop (RXL) are loop coils consisting of a single turn (shown in \mbox{Fig. \ref{fig_loop}}), the former connecting to a voltage source, V\textsubscript{source}, with a source impedance, R\textsubscript{source}, and the latter connecting to a load, R\textsubscript{load}. L\textsubscript{1}, L\textsubscript{4} and R\textsubscript{p1}, R\textsubscript{p4} designate self-inductance and parasitic resistances of the two loop coils respectively, and C\textsubscript{1}, C\textsubscript{4} designate the tuned capacitance, which set the loop coils to resonate at the designed frequency. The TX Coil (TXC), IC, and RX Coil (RXC) are multi-turn single layer spiral coils that have self-inductance of L\textsubscript{2}, L\textsubscript{i},  L\textsubscript{3} and parasitic resistance of R\textsubscript{p2}, R\textsubscript{pi}, R\textsubscript{p3}, respectively. Capacitors C\textsubscript{2}, C\textsubscript{i}, C\textsubscript{3} are added to tune the spiral coils to the resonant frequency. M\textsubscript{12} and M\textsubscript{34} are the mutual inductance (M) between TXL and TXC and between the RXL and RXC, respectively. M\textsubscript{2i} and M\textsubscript{3i} are the M between TXC and IC and between the RXC and IC. The M of nonadjacent elements are neglected, e.g. M\textsubscript{14}, M\textsubscript{13}, etc., as they are minuscule by comparison \cite{article5}. 

\begin{align} \label{coil_geometry}
    \begin{aligned}
        ID_{spiral} &= 2r_{1} \\
        OD_{spiral} &= 2r_{N}
    \end{aligned}
\end{align}

The generalized self-inductance L is defined by the geometry of a coil. As shown in \mbox{Fig. \ref{fig_coilgeometry}}, parameters of both loop coil and multi-turn single layer spiral coil are defined. In \mbox{Fig. \ref{fig_loop}}, OD\textsubscript{loop} and w indicate the OD of the loop coil and the wire diameter respectively. In \mbox{Fig. \ref{fig_spiral}}, w, N, p, and R\textsubscript{N} indicate the wire diameter, the number of turns, distance between each turn, and the radius of N\textsuperscript{th} turn respectively. The inner diameter, ID\textsubscript{spiral}, and the outer diameter, OD\textsubscript{spiral}, of a spiral coil is defined in (\ref{coil_geometry}). All units of length are in meters. 

The equation for the L\textsubscript{x} of spiral coils (\ref{spiral_self_inductance}) is defined in \cite{inproceedings3}, which is a modified form of Wheeler's formula. For loop coils (\ref{loop_self_inductance}), it's a modified form of Kirchhoff's formula which can be found in \cite{inbook1}. In both, the inductance is measured in Henries. 

\begin{subequations}
    \begin{align} \label{spiral_self_inductance}
        L_{spiral} = \frac{N^{2}(OD_{spiral}-N(w+p)^{2})}{16OD_{spiral}+28N(w+p)}\frac{39.37}{10^{6}}
    \end{align}
    \begin{align} \label{loop_self_inductance}
        L_{loop} &= 2\pi OD_{loop}(\log \frac{4OD_{loop}}{w}-1.75)10^{-7}
    \end{align}
\end{subequations}

The generalized equation for M\textsubscript{AB} from \cite{inproceedings3} is modified by summing all the mutual inductances of i\textsuperscript{th} turn of the primary coil A to j\textsuperscript{th} turn of the secondary coil B \ref{mutual_inductance}, and has units of Henries. The radii r\textsubscript{Ai} and r\textsubscript{Bj}, N\textsubscript{A} and N\textsubscript{B} define the geometry of the two coils, d indicates the distance between them, and $\mu_{0}$ is the free space permeability. 

\begin{align} \label{mutual_inductance}
    \begin{split}
        M_{AB} &= \sum ^{N_{A}}_{i=1}\sum ^{N_{B}}_{j=1}\mu_{0}r_{Ai}r_{Bj} \\
        & \int ^{\pi }_{0}\frac{\cos{\theta}}{\sqrt{r_{Ai}^{2}+r_{Bj}^{2}+d^{2}-2r_{Ai}r_{Bj}\cos{\theta}}}d\theta
    \end{split}
\end{align}

The coupling coefficient \cite{article5} is proportional to the mutual inductance between the primary and the secondary coils (\ref{coupling_coefficient0}), normalized by the self-inductances. 

\begin{align} \label{coupling_coefficient0}
    k_{AB} = \frac{M_{AB}}{\sqrt{L_AL_B}}
\end{align}

The tuned capacitance (\ref{tuned_capacitor}) is defined by the equation of resonant frequency of RLC circuits and its value is assumed to be dominant as it is generally large enough that the self-capacitance of a coil is negligible \cite{inproceedings3} addition. The f\textsubscript{0} in the equation indicates the designed resonant frequency, and the unit of capacitance is in Farad.
\begin{align} \label{tuned_capacitor}
    C = \frac{1}{(2\pi f_{0})^{2}L}
\end{align}

The parasitic resistance of a coil, R\textsubscript{p}~(\ref{conduction_loss}), consists of both a conductive as well as radiative loss\cite{inproceedings4}. The frequency dependent conduction loss at high frequencies is better described as the resistive skin effect, R\textsubscript{skin}~(\ref{skin_effect_loss}), which modifies  R\textsubscript{DC} (\ref{DC_loss}) to account for shell distribution of current within in a wire at high frequencies. The radiative proximity effect loss, R\textsubscript{proximity}, results from the electron crowding due to EM fields of nearby wires within a given coil (\ref{proximity_effect_loss}).

\begin{subequations}
    \begin{align} \label{conduction_loss}
        R_p = R_{skin} + R_{proximity}
    \end{align}
    \begin{align} \label{DC_loss}
        R_{DC}=\frac{l}{\sigma\pi(\frac{w}{2})^2},~l = \frac{1}{2}N\pi(OD+ID)
    \end{align}
    \begin{align} \label{skin_effect_loss}
        \begin{split}
            R_{skin} = R_{DC}\frac{\gamma}{2}\frac{ber(\gamma)bei^{'}(\gamma) - bei(\gamma)ber^{'}(\gamma)}{(ber^{'}(\gamma))^2 + (bei^{'}(\gamma))^2} \\
            \gamma = \frac{w}{\delta\sqrt{2}}, ~\delta=\frac{1}{\sqrt{\pi f\sigma\mu_{0}}}
        \end{split}
    \end{align}
    \begin{align} \label{proximity_effect_loss}
        R_{proximity} = R_{DC}\frac{-2\pi\gamma}{2}\frac{ber_2(\gamma)ber^{'}(\gamma)+bei_2(\gamma)bei^{'}(\gamma)}{(ber(\gamma))^2+(bei(\gamma))^2}
    \end{align}
\end{subequations}

Where $\sigma$ is the copper's conductivity, $\delta$ denotes the skin depth.

By applying Kirchhoff's Voltage Law (KVL), the relationship between current and voltage in each coil is shown in (\ref{matrix}). 

\begin{align} \label{matrix}
    \begin{vmatrix}
    I_{1} \\
    I_{2} \\
    I_{i} \\
    I_{3} \\
    I_{4}
    \end{vmatrix}
    \begin{vmatrix}
    Z_{11}  & Z_{22}    & 0         & 0         & 0     \\
    Z_{21}  & Z_{22}    & Z_{2i}    & 0         & 0     \\
    0       & Z_{i2}    & Z_{ii}    & Z_{i3}    & 0     \\
    0       & 0         & Z_{3i}    & Z_{33}    & Z_{34}\\
    0       & 0         & 0         & Z_{43}    & Z_{44}
    \end{vmatrix}=
    \begin{vmatrix}
    V_{source}  \\
    0           \\
    0           \\
    0           \\
    0
    \end{vmatrix}
\end{align}

Then the ratio of load to source voltage provides the voltage gain, as solved from (\ref{matrix}), and yields (\ref{vload_over_vsource}).

\begin{align} \label{vload_over_vsource}
    \frac{V_{load}}{V_{source}}=\frac{Z_{12}Z_{2i}Z_{3i}Z_{34}R_{load}}{\left(
        \splitfrac{Z_{11}Z_{22}Z_{ii}Z_{33}Z_{44}+Z_{11}Z_{2i}^{2}Z_{34}^{2}}{
            \splitfrac{+Z_{ii}Z_{12}^{2}Z_{34}^{2}+Z_{44}Z_{12}^{2}Z_{3i}^{2}}{
                \splitfrac{-Z_{12}^{2}Z_{ii}Z_{33}Z_{44}-Z_{2i}^{2}Z_{11}Z_{33}Z_{44}}{
                    -Z_{3i}^{2}Z_{11}Z_{22}Z_{44}-Z_{34}^{2}Z_{11}Z_{22}Z_{ii}
                }
            }
        }
    \right)}
\end{align}

The impedances along the diagonal in (\ref{matrix}), e.g. Z\textsubscript{11} to Z\textsubscript{44}, represent the self-impedance of each coil, which consist of both resistance and reactance as detailed in (\ref{self_impedance}). 

\begin{align} \label{self_impedance}
    \begin{aligned}
        Z_{11} &= R_{source}+R_{p1}+j\omega L_{1}+\frac{1}{j\omega C_{1}} \\ 
        Z_{22} &= R_{p2}+j\omega L_{2}+\frac{1}{j\omega C_{2}} \\
        Z_{ii} &= R_{pi}+j\omega L_{i}+\frac{1}{j\omega C_{i}} \\
        Z_{33} &= R_{p3}+j\omega L_{3}+\frac{1}{j\omega C_{3}} \\
        Z_{44} &= R_{load}+R_{p4}+j\omega L_{4}+\frac{1}{j\omega C_{4}}
    \end{aligned}
\end{align}

By contrast the off-diagonal elements of  (\ref{vload_over_vsource}), e.g. Z\textsubscript{12} or Z\textsubscript{2i}, are the mutual-impedance of two adjacent coils. They can be substituted for M\textsubscript{AB} via (\ref{mutual_impedance}) \cite{article12}.

\begin{align} \label{mutual_impedance}
    \begin{aligned}
        Z_{12} &= Z_{21} = j\omega M_{12} \\ 
        Z_{2i} &= Z_{i2} = j\omega M_{2i} \\ 
        Z_{3i} &= Z_{i3} = j\omega M_{3i} \\ 
        Z_{34} &= Z_{43} = j\omega M_{34} 
    \end{aligned}
\end{align}

Finally, the scattering parameter S\textsubscript{21} (\ref{S21}) can be can be derived from (\ref{vload_over_vsource}) by using \cite{article5}. Note that the S\textsubscript{21} generally refers to the power received by port 2 from port 1, which is the power received by load from source in our system, and is thus designated here in order to stick to accepted nomenclature. 

\begin{align} \label{S21}
    S_{21} &= 2\frac{V_{load}}{V_{source}}\sqrt{\frac{R_{source}}{R_{load}}}
\end{align}

\begin{figure}[!b]\centering
	\includegraphics[width=0.48\textwidth]{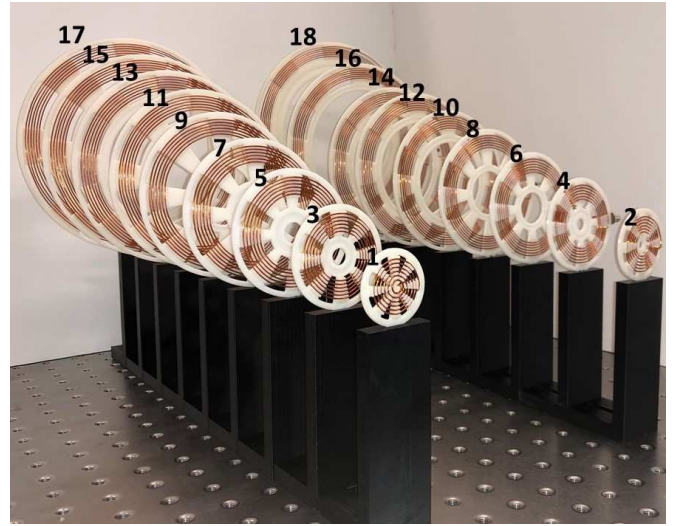}
	\caption{The tested ICs range from OD = 30mm to 200mm, in 10mm increments. 
	}
	\label{fig_intermediate_coils}
\end{figure}

\begin{figure}[!t]\centering
    \includegraphics[width=\linewidth]{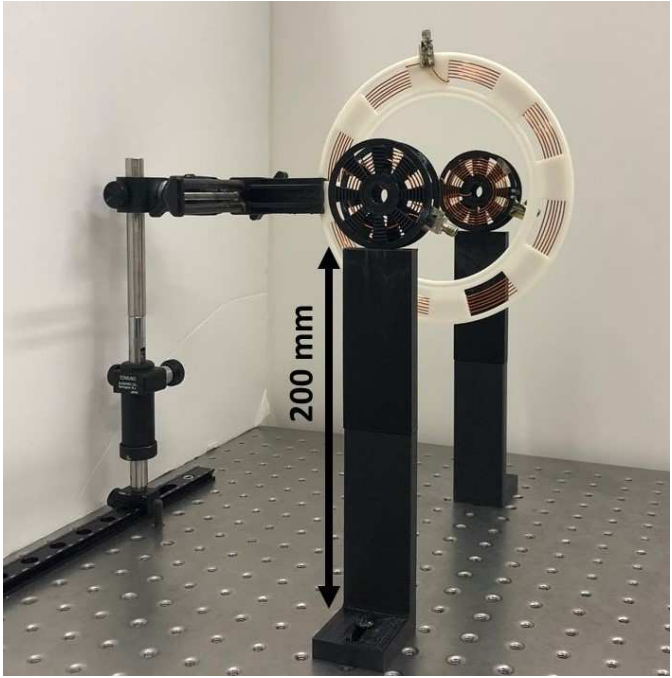}
    \caption{A rotated view of the experimental setup showing (from foreground to background) a given instance of the TX, IC, and RX variations.}
    \label{fig_experiment_setup}
\end{figure}

\section{System Setup} \label{system_setup}

\subsection{Experimental Setup} \label{experimental_setup}
To minimize sources of variation, all of the coils have been designed to have exactly seven turns, using the same wire diameter of the same wire type of wire (Magnet wire, 20 AWG), and having the same distance between each turn. Therefore, the primary independent variable for each coil is their OD. \mbox{Fig. \ref{fig_five_coil_system}} is an aerial view for both symmetric and asymmetric systems, and all loops and coils are axially aligned. The separation between TXL and TXC, as well as RXL and RXC, is held at the distance where S\textsubscript{11} is minimized (symmetric separation = 2.5mm, whereas in the asymmetric system, the large coil's separation = 10mm and the small coil's separation = 2mm). Using the above biased TX and RX in a series resonant circuit allows for the doubling of the coupling coefficient \cite{article9}, and therefore is advantageous when attempting to maximize transferred power over longer separations between TX and RX. 

The distance between the face of TX and RX, and therefore also TXC and RXC, is a constant 150mm. We sweep the IC from one end to the other, and its location along the axis is denoted by the separation between the TX and the IC. All coils are 200 mm above the metal experimental platform so that it will not interfere the WPT system. As show in \mbox{Fig. \ref{fig_intermediate_coils}}, we had designed ICs with ODs from 30 to 200 mm at 10 mm interval. \mbox{Fig. \ref{fig_experiment_setup}} shows the experimental set-up of the symmetric system where the TX and the RX are identical. In this symmetric system, as a reference for this work, both the TXC and RXC are designed to be multi-turn single layer spiral coils that have a 50 mm outer diameter. Both TXL and RXL are loop coils that have a 38 mm OD. In the case of the asymmetric system, we have designed the OD of the TXC is 90 mm, that of the RXC is 30 mm, that of the TXL is 67.5 mm, and that of the RXL is 10.44 mm. The OD of the spiral coils to be tested was chosen to ensure the TX and RX are weakly coupled. Thus, the OD of TX was chosen to have a critical coupling distance of around 1/5 the TX to RX separation (150mm), where the critical coupling distance is $d_{TX-RX}= r_{N}/\sqrt{2}$ \cite{article13}. Then, the OD of RSC was chosen to be the minimum size possible while maintaining the same coil geometric associations (i.e. wire diameter, number of turns, and the separation between each turn). The scattering parameter S\textsubscript{21} is measured by the HP8753E Vector Network Analyzer (VNA) at the designed resonant frequency. 

\subsection{Quantitative Simulation Setup} \label{quantitative_simulation_setup}
To provide additional insight towards optimal IC location and size selection, the chain of analysis followed in Section~\ref{system_analysis} was replicated in Matlab to mimic the experimental setup in \ref{experimental_setup}. Key parameters, such as OD, number of turns, wire diameter, and separation between each turn, have the same values of those in the experimental setup. Within the simulation, the self-inductance (\ref{spiral_self_inductance}) (\ref{loop_self_inductance}) and mutual inductance (\ref{mutual_inductance}) are then calculated based on a given coils' geometry. In order to tune all coils to the designated resonant frequency, the value of tuning capacitance is derived from (\ref{tuned_capacitor}). Finally, from self-impedance (\ref{self_impedance}) and mutual impedance (\ref{mutual_impedance}) the magnitude of S\textsubscript{21} (\ref{S21}) is derived and simulation results are reported alongside the experimental. 

\section{Experimental and Quantitative Simulation Results} \label{experimental_and_simulation_results}

\subsection{Symmetric System}
\begin{figure}[!ht]\centering 
    \includegraphics[width=\linewidth]{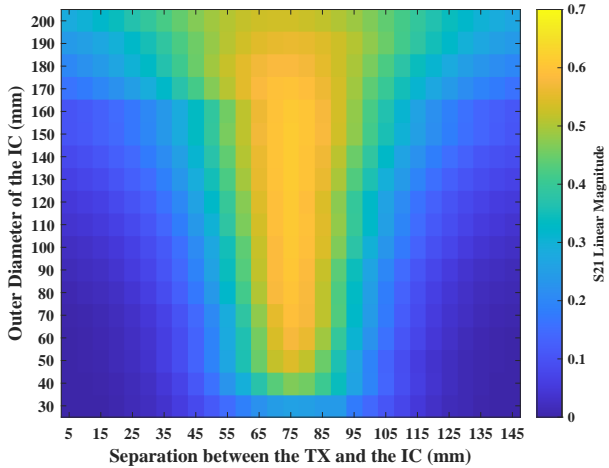}
    \caption{In the symmetric system (TXC OD = RXC OD = 50mm), a S\textsubscript{21} heatmap resulting from sweeping ICs of different sizes(OD = 30mm to 200mm) through the separation between the TX and RX.}
    \label{fig_symmetric_intermediate}
\end{figure}

In the symmetric system, \mbox{Fig. \ref{fig_symmetric_intermediate}} results from sweeping the ICs of different sizes along the axis between the TX and RX. The horizontal axis shows the separation between the TX and the IC. The vertical axis shows the OD of the IC of different sizes. The intensity of color represents the magnitude of S\textsubscript{21}. The maximum magnitude of S\textsubscript{21} within a given separation sweep is always located where the separation between the TX and the IC equals 75mm, which is the center of the TX and RX in the symmetric system, and the coupling coefficient of the TXC and the IC, k\textsubscript{TXC-IC}, equals the coupling coefficient of the RX and the IC, k\textsubscript{RXC-IC}. Our simulation shows the maximum magnitude of S\textsubscript{21} and equal-coupling points are both located where the IC is centered between the TX and the RX.

\subsection{Asymmetric System}
\begin{figure}[!ht]\centering 
    \includegraphics[width=\linewidth]{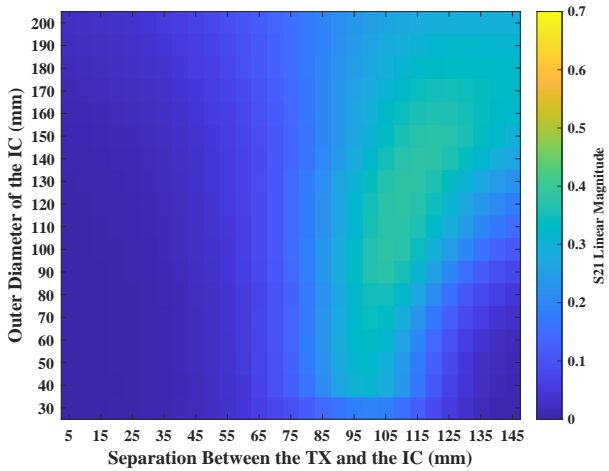}
    \caption{In the asymmetric system (TXC OD = 90mm, RXC OD = 30mm), a S\textsubscript{21} heatmap resulting from sweeping ICs of different sizes (OD = 30 mm to 200mm) through the separation between the TX and RX.}
    \label{fig_asymmetric_intermediate}
\end{figure}

Similar to the symmetric system, \mbox{Fig. \ref{fig_asymmetric_intermediate}} is the result of sweeping ICs of different sizes in the asymmetric system. The maximum magnitude of S\textsubscript{21} is always closer to the RX, which is the smaller coil side. This phenomenon has also been analyzed and experimentally shown in \cite{article9}. 
In addition, the location of ICs where maximum S\textsubscript{21} present in the asymmetric system gradually shifts closer to the RX as the OD of ICS gets larger. Note, this result is mirrored when RX replaces TX as the the transmitter (i.e. S\textsubscript{21} = S\textsubscript{12}).

\begin{figure}[!ht]\centering 
    \includegraphics[width=\linewidth]{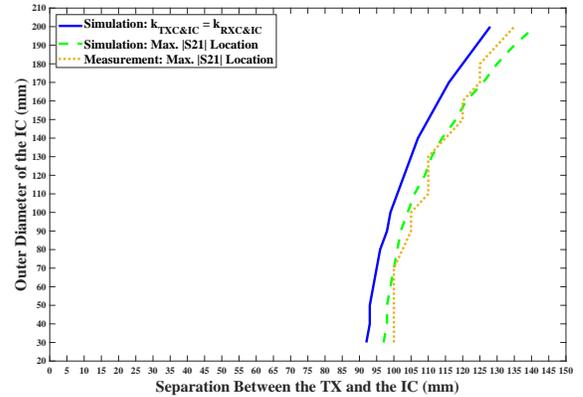}
    \caption{In the asymmetric system (TXC OD = 90mm, RXC OD = 30mm), Equal-coupling location (Simulation) (mm) vs. Maximum S\textsubscript{21} Location (Simulation) (mm) vs. Maximum S\textsubscript{21} Location (Measurement) have been compared through the separation between the TX and RX.}
    \label{fig_asymmetric_simulation_vs_measurement}
\end{figure}

 As shown in \mbox{Fig \ref{fig_asymmetric_simulation_vs_measurement}}, we compare the location of the k\textsubscript{TXC-IC} = k\textsubscript{RXC-IC} in simulation and the location of the IC where the maximum magnitude of S\textsubscript{21} present in simulation and measurement. The simulated maximum of S\textsubscript{21} location matches with the measured maximum of S\textsubscript{21} location in general, and the simulated equal-coupling location is different from the measurement by 6.67\% in most of the tested cases because we have neglected the inter-coupling coefficients (k\textsubscript{TXL-RXC}, k\textsubscript{TXL-RXL}, etc.).

\subsection{Optimal IC Size}
While it may seem that a larger IC may result in monotonically increasing magnitude of S\textsubscript{21},  \mbox{Fig. \ref{fig_symmetric_intermediate}} and \mbox{Fig. \ref{fig_asymmetric_intermediate}} show experimental results that suggest the optimal size of an IC in both symmetric and asymmetric systems. In \mbox{Fig. \ref{fig_symmetric_intermediate}} and \mbox{Fig. \ref{fig_asymmetric_intermediate}}, the magnitude of S\textsubscript{21} will increase with increasing size of ICs up to a given OD, but then continuously decrease after it is larger than a certain size. With the given experimental setup, the optimal size of the IC is OD 150mm for the symmetric system, which is 3x larger than the OD 50mm TXC and RXC, and 140 mm for the asymmetric system, which is 1.6x larger than the OD 90mm TXC and 4.7x larger than the OD 30mm RXC. 

\section{Conclusion}

After analyzing the circuit model of the five-coil system, and through the above experiments, quantitatively simulated and experimentally measured results have been compared. We have found that the optimal location of an IC in both symmetric and asymmetric systems is located where the TX and RX are equally coupled to the IC. Then, from the experimental results in the given setup, we have determined that there exists an optimal size of IC in both symmetric system and asymmetric system, and the size is larger than both TX and RX.

Future work includes deriving more explicit formulas for better predicting the optimal size in the above 5-coil setup, and incorporating an optimal separation of the loop and spiral coils for both the TX and RX.

\section*{Acknowledgment}
This work was supported by Department of Defense Medical Command Award No. W81XWH-20-1-0010 and the Center for Neural Technology (CNT), NSF Grant EEC-1028725, as well as University of Washington Institute for Neuroengineering (UWIN). The authors would also like to specifically thank the members of the Sensor Systems Lab for their help and support in this work, and Han-Ting Lin for his support.



\begin{IEEEbiography}[{\includegraphics[width=1in,height=1.25in,clip,keepaspectratio]{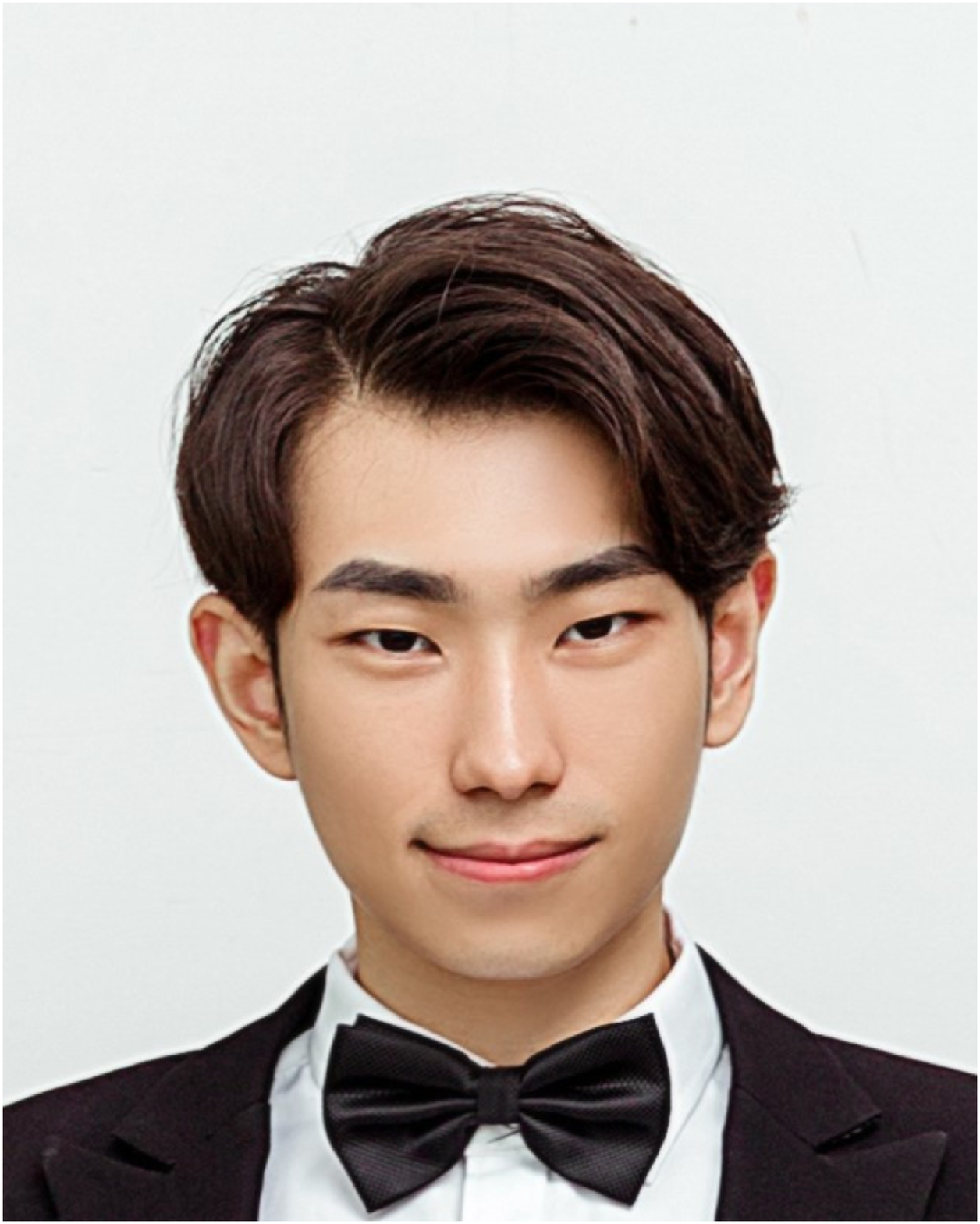}}]
{Kedi Yan} (S'20) was born in Zhengzhou, Henan, China on August 22nd, 1995. He received the B.S. degree in Electrical and Computer Engineering from Oregon State University, Corvallis, OR, USA in 2018. He is currently working toward his M.S. degree in Electrical Engineering at the University of Washington, Seattle, WA, the United State. 

His research interests include electromagnetism in general, with a special focus on Radio Frequency, Microwave, WPT, WPT to biomedical devices, and Magnetic Nanoparticles.

\end{IEEEbiography}

\begin{IEEEbiography}[{\includegraphics[width=1in,height=1.25in,clip,keepaspectratio]{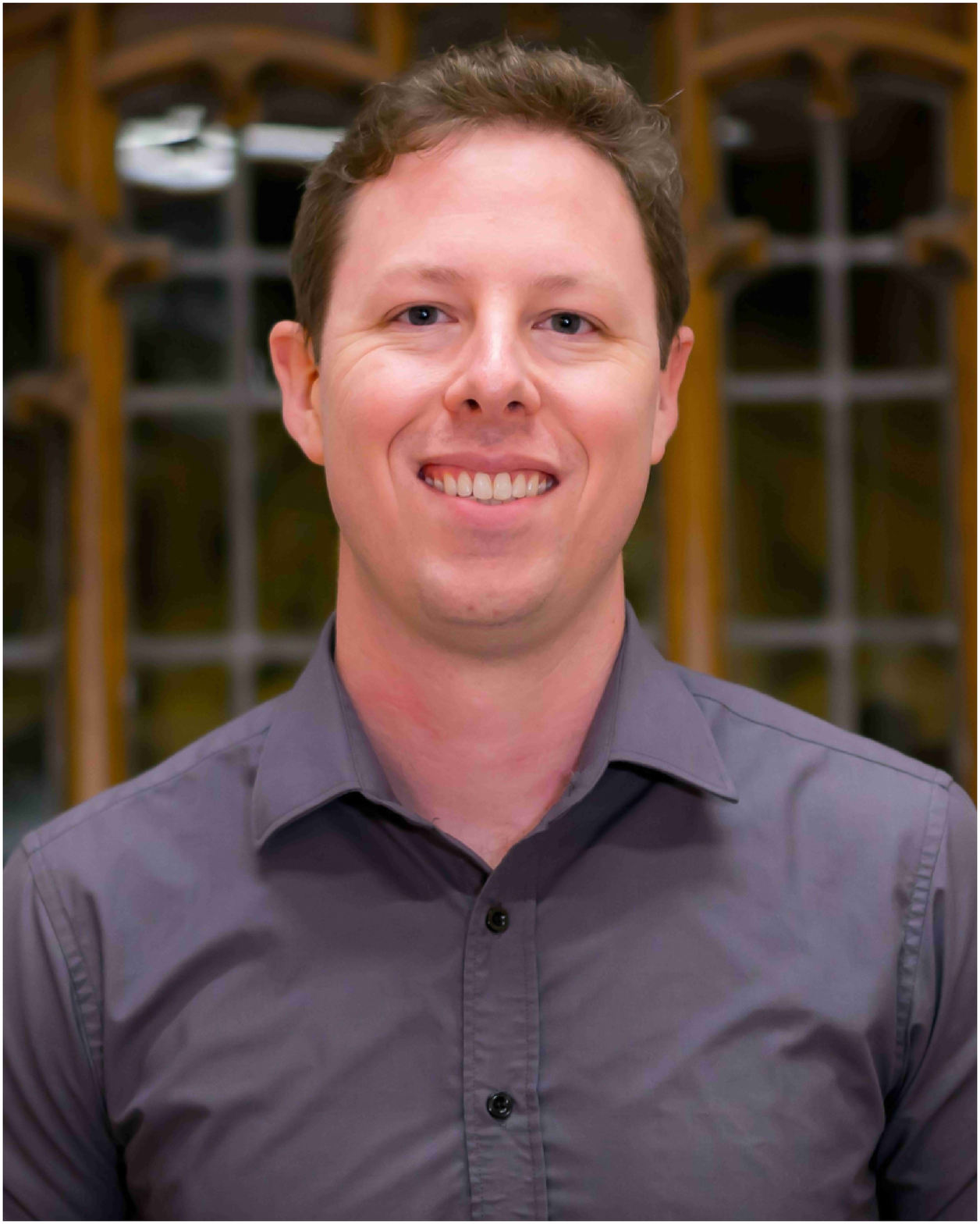}}]
{Gregory Moore} (M'20) was born in New Orleans, LA, USA. He received his B.S. degree in Physics and B.S. degree in Electrical Engineering from the University of Maryland, College Park, MD, USA in 2005, and the M.S. degree in Electrical Engineering from the University of Washington, Seattle, WA, USA in 2018.

From 2008 to 2012 he was with Tao of Systems Integration, Inc., Hampton, VA, USA, where he was engaged mixed signal circuit and control systems design for flow measurement products for the aeronautics/marine engineering. He is currently a Research Assistant at the University of Washington, WA, USA. His research is concerned with WPT and low-power communication design, with special focus on applications directed towards biopotential stimulation, recording, and telemetry.
\end{IEEEbiography}

\begin{IEEEbiography}[{\includegraphics[width=1in,height=1.25in,clip,keepaspectratio]{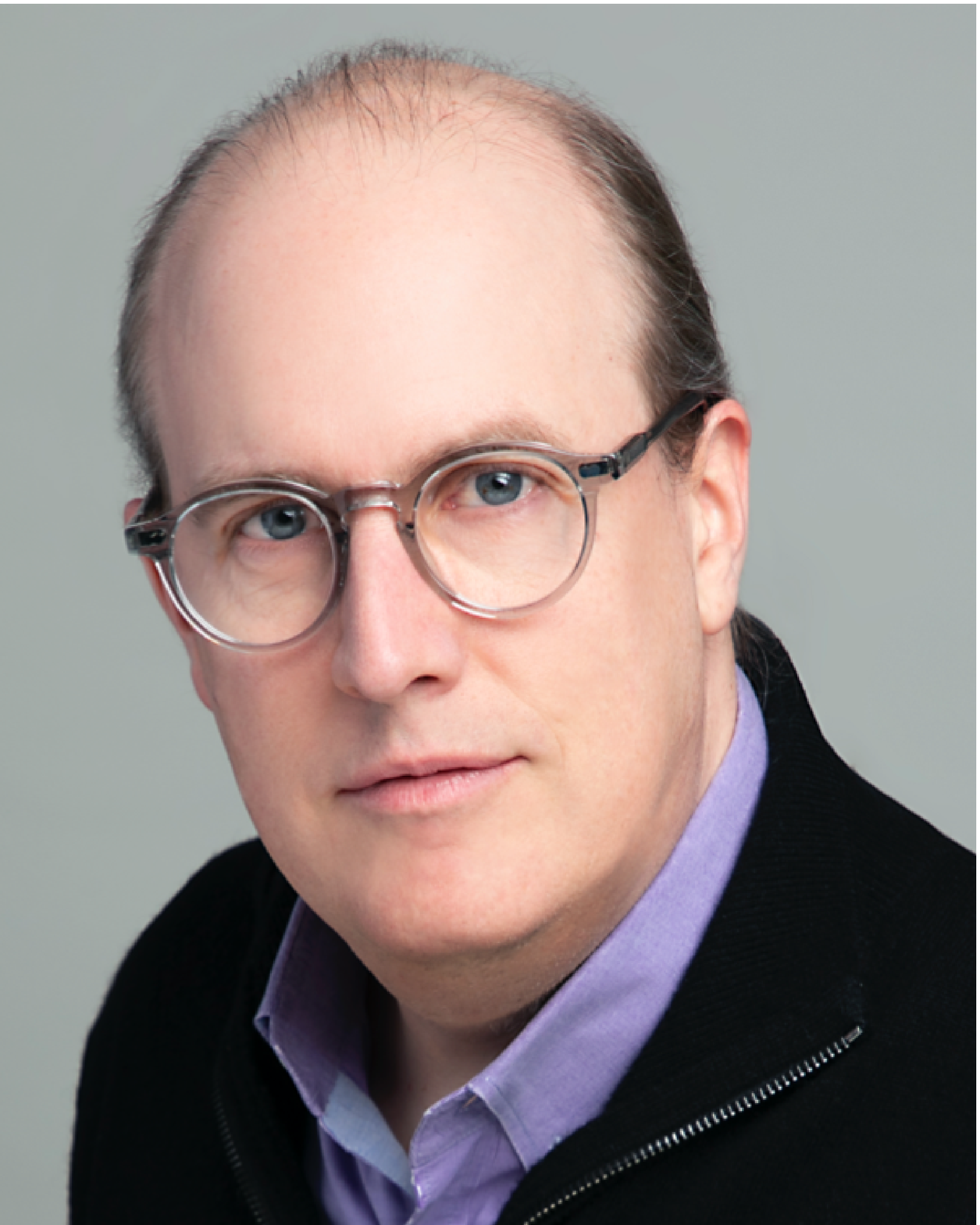}}]
{Joshua R. Smith} (M'99-SM'12-F'20) was born in Washington, D.C.. He received a B.A. in Computer Science and Philosophy from Williams College, Williamstown, MA, USA in 1991, an M.A. in Physics from the University of Cambridge, Cambridge, UK in 1997, and S.M. and Ph.D. degrees from the Media Laboratory at the Massachusetts Institute of Technology, Cambridge, MA, USA in 1995 and 1999.

He is the Milton and Delia Zeutschel Professor, jointly appointed in Computer Science and Engineering and in Electrical and Computer Engineering at the University of Washington. He is a founder of three companies: Wibotic, Jeeva Wireless, and Proprio. Previously he was a Principal Engineer at Intel Corp.

Professor Smith was elevated to IEEE Fellow for contributions to far- and near-field wireless power, backscatter communication, and electric field sensing. He is interested in all aspects of sensor systems, including wireless methods for powering and communicating with sensors, and applications including ubiquitous computing, biomedical implants, and robotics.

\end{IEEEbiography}

\end{document}